\documentclass[epsfig,psfig]{aa}

\usepackage{graphicx}
\usepackage{txfonts}
\begin{document}

\title{GalICS II: the [$\alpha$/Fe]-mass relation in elliptical galaxies}
\titlerunning{GalICS II: [$\alpha$/Fe] ratios in ellipticals}

\author{Antonio Pipino$^{1,2}$, Julien E.G.Devriendt$^{1,3}$, Daniel Thomas$^4$, Joseph Silk$^{1}$ \& Sugata Kaviraj$^{1}$}
\authorrunning{A. Pipino et al.}

\institute{$^1$ Astrophysics, University of Oxford, Denys Wilkinson Building, Keble Road, Oxford OX1 3RH, UK\\
$^2$ Department of Physics and Astronomy, University of Southern California, Los Angeles,CA 90089-0484 \\
$^3$ Observatoire Astronomique de Lyon, 9 avenue Charles Andre, 69561 Saint Genis Laval Cedex, France\\
$^4$ Institute of Cosmology and Gravitation, University of Portsmouth, Mercantile House, Hampshire Terrace, Portsmouth, PO1 2EG, UK}
\date{Accepted,
      Received }

%\maketitle

\abstract{} 
{We aim at reproducing the mass- and $\sigma$-[$\alpha$/Fe]
relations in the stellar populations of early-type galaxies by means of a
cosmologically motivated assembly history for the spheroids.}  
{We implement a detailed treatment for the chemical evolution of H, He, O
and Fe in GalICS, a semi-analytical model for galaxy formation which
successfully reproduces basic low- and high-redshift galaxy properties.
The contribution of supernovae (both type
Ia and II) as well as low- and intermediate-mass stars to 
chemical feedback are taken into account. The model predictions are
compared to the most recent observational results.}  
{We find that this chemically improved GalICS does not produce the observed mass- and $\sigma$-[$\alpha$/Fe] relations. The slope is too shallow and scatter too large, in particular in the low and intermediate mass range. The model shows significant improvement at the highest masses and velocity dispersions, where the predicted [$\alpha$/Fe] ratios are now marginally consistent with observed values. We show that this result comes from the implementation of AGN (plus halo) quenching of the star formation in massive haloes. A thorough exploration of the parameter space shows
that the failure of reproducing the mass- and $\sigma$-[$\alpha$/Fe] relations can partly be attributed to the way in which star formation and feedback are currently modelled. The merger process is responsible for a part of the scatter. We suggest that the next generation of semi-analytical model should feature feedback (either stellar of from AGN) mechanisms linked to single galaxies and not only to the halo, especially in the low and intermediate mass range.}  
{The integral star formation history of a single galaxy determines its
final stellar [$\alpha$/Fe] as it might be expected from the results of closed
box chemical evolution models. However, the presence of dry-mergers and
metal recycling in the hot gas phase helps in keeping the $\alpha$ element
abundance in the stars at a super-solar level in a hierarchical galaxy
formation scenario.}


\maketitle

\section{Introduction}

The Cold Dark Matter (CDM) scenario (Peebles, 1982) successfully explains the
growth of the large scale structure of the universe (Springel, Frank \& White 2006). Since the
original application of this scenario to the galaxy formation process
problem (Kauffmann \& White, 1993, Cole et al 1994), however, several modifications to the pure hierarchical
assembly of the building blocks had to be introduced in order to deal
with the complexity of baryonic physics.  Among the main open issues,
we mention the anti-hierarchical behaviour of the AGNs (e.g. Hasinger
et al. 2005), the evolution of luminosity function with red-shift
(e.g. Bundy et al. 2005) as well as the
%DTstart
increase of mean stellar [$\alpha$/Fe] with galaxy mass (or
$\sigma$) in elliptical galaxies (e.g., Worthey et al 1992; Trager et al 2000; Thomas et al 2005, Nelan et al. 2005).%cite
This relationship together with the old inferred ages implies that more massive ellipticals formed earlier and faster with respect to smaller objects (Matteucci 1994; Thomas et al. 2005).  
In the following we will refer to the above observational constraints with the term \emph{downsizing} (Cowie et al. 1996).

Thomas (1999) was the first to study the chemical enrichment of $\alpha$ and Fe-peak elements in the framework of hierarchical models of galaxy formation. In a very simplistic approach, Thomas (1999)%cite 
ran chemical evolution simulations over the star formation histories predicted by Kauffmann (1996)
neglecting the complex merger history of the galaxies. It turned out that the predicted star formation histories of massive elliptical galaxies were to extended to produce $\alpha$/Fe consistent with observations. This conclusion was later reinforced by Nagashima et al. (2005), who included a self-consistent treatment of chemical enrichment in semi-analytic galaxy formation models, and by Pipino \& Matteucci (2006, PM06).

More recently, a plethora of new models have been presented to address the general phenomenon of downsizing (e.g. Croton et al 2006, De Lucia et al. 2006, Bower et al 2006, Cattaneo et al., 2005, Sommerville et al. 2008, Kaviraj et al. 2005, Fontanot et al., 2007).
The new key ingredient in these models is feedback from super-massive black holes, which is used to suppress residual star formation at late times in the evolution of massive galaxies (Granato et al. 2004). Such a suggested scenario seems to be supported by observations (e.g. Nesvadba et al. 2006).
In practice, the mass assembly still occurs at late times in these models , whereas most of the stars have been formed at high red-shift in small sub-units
(but see Cimatti et al. 2006). The preferred mechanism for the assembly
of massive spheroids is a sequence of dissipation-less (\emph{dry})
mergers. This leads generally to better agreement with the observed downsizing pattern.

However, the [$\alpha$/Fe]-mass relation has not been studied in these new generation models (see Pipino \& Matteucci 2008, PM08). The aim of this paper is to fill this gap. 
%DTend
To this end we implement a fully self-consistent treatment of the chemical
evolution, which includes a robust estimate of the type Ia supernova
rate and of the Fe production into GalICS (Hatton et al., 2003; Paper
I hereafter), a state-of-the-art semi-analytical model for galaxy formation and
evolution based on a CDM-driven growth of the
structures.  The main goal is to check model predictions against the latest
observational results for the $\alpha/Fe$-mass relation and the mass-metallicity relation (MMR, e.g. Carollo
et al. 1993), that so far could be simultaneously accounted for only by revised monolithic models (Pipino \& Matteucci, 2004, PM04).
The new results will be then interpreted in the light of our
previous work with the chemical evolution models (Thomas 1999, PM04, PM06, PM08).
%% The main novelty of this paper is the
%% extensive study of the physical quantities which affect slope,
%% zero-point and scatter of the predicted $\alpha/Fe$-mass relation and MMR in order to assess which
%% is the main driver of these relations.

The structure of the paper is as follows:
the main improvements with respect to Paper I are
spread out in Sec.2; in Sec.3 the chemical evolution scheme
is tested against the Milky Way and the local SNIa rate.
In Sec. 4, 5 and 6 the results are presented and discussed,
respectively.

\section{The model}

\subsection{The GalICS galaxy formation model}

GalICS is a model of hierarchical galaxy formation which combines high
resolution cosmological N-body simulations to describe the dark matter
content of the Universe with semi-analytic prescriptions to follow the
physics of the baryonic matter.  GalICS has been thoroughly presented in Paper
I, which we refer the reader to for a detailed discussion of the model
assumptions and properties. It has already been used for the study
of the colour-magnitude relation and the progenitor bias of elliptical galaxies (Kaviraj et al. 2005), 
the reproduction of the Galex NUV-optical colours (Kaviraj et al. 2007) and
the black-hole mass - $\sigma$ relation (Cattaneo et al 2005). 
It has also been used to explore the consequences the halo-quenching mechanism 
(Keres et al. 2005) by Cattaneo et al. (2008 and references
therein).  
The above mentioned papers represent a comprehensive set of benchmark tests that we will not 
repeat here, but simply point out that results are preserved to a large extent in our
present implementation.

We briefly recall the specifications of cosmological N-body simulation
used to construct the halo merger trees.  This simulation is a realization of 
flat cold dark matter universe with a cosmological constant of
$\Omega_{\Lambda}=0.667$.  The simulated volume is a cube of side
$L_{\rm box}=100h_{100}^{-1}$Mpc, with $h_{100}\equiv
H_0/100{\rm\,km\,s}^{-1}=0.667$, which contains $256^3$ particles of mass
$8.3\times 10^9$M$_{\odot}$ each, the cold dark matter power spectrum
was normalised in agreement with the present day abundance of rich
clusters ($\sigma_8 =0.88$).  
One should bear in mind that the dark
matter simulation cannot resolve haloes less massive than $1.6\times
10^{11}M_\odot$, which implies that  a galaxy less massive than $2\times
10^{10}M_\odot$ is formally below the resolution limit. 
The spatial resolution, instead, is such that we cannot resolve
scales below 30 kpc.

As hot gas cools and falls to the centre of its dark matter halo, 
it settles in a rotationally supported disc.  According to Paper I, if the specific 
angular momentum of the accreted gas is conserved and starts off with 
the specific angular momentum of the dark matter halo, 
we assume it forms an exponential disc with scale length $r_d$ given by:
\begin{equation}
\label{eqn:disc_size}
r_d = \frac{\lambda}{\sqrt{2}} R_{200}.
\end{equation}

Galaxies remain pure discs if their disc is globally 
stable ({\em i.e.} $V_c < 0.7 \times V_{tot}$ where $V_{tot}$ is the 
circular velocity of the disk-bulge-halo system ; see e.g. van den Bosh et al 1998), 
and they do not undergo a merger with another galaxy. 
In the case where the latter of these two events occurs, 
we employ a recipe to distribute the stars 
and gas in the galaxy between three components in the resulting, 
post-merger galaxy, that is the disc, the bulge, and a star-burst (see Paper I).  
In the case of a disc instability, we simply transfer the mass of gas and stars 
necessary to make the disc stable to the burst component, and compute the properties of 
the bulge/burst in a similar fashion as that described in Paper I.
Bulges are assumed to have a density profile given by Hernquist (1990).
The bulges are assumed to be pressure supported with a characteristic 
velocity dispersion $\sigma$, computed at their half-mass radius.  

Galaxy morphology in the model is determined by the ratio of the
B-band luminosities of the disc and bulge components. A morphology
index is defined as

\begin{equation}
I = \exp\bigg(\frac{-L_B}{L_D}\bigg)
\end{equation}

such that a pure disc has $I=1$ and a pure bulge has $I=0$.
Following Baugh et al (1996), ellipticals have $I<0.219$, S0s have
$0.219<I<0.507$ and spirals have $I>0.507$. This simple
prescription is clearly incapable of capturing the complex
spectrum of real galaxy morphologies. Therefore, in what follows,
`spirals' refer to \emph{all} systems which do not have a dominant
spheroidal (bulge) component. Observationally, this includes not
only systems with distinctive spiral morphologies, but also
peculiar or irregular systems.

\subsection{The chemical evolution}

The main novelty of the present versions of GalICS is the
implementation of a self-consistent treatment of the chemical
evolution with finite stellar lifetimes and both type Ia and type II
supernovae ejecta.  In practice, we follow the chemical evolution of
only four elements, namely H, He, O and Fe. This set of elements is
good enough to characterise our simulated elliptical galaxy from the
chemical evolution point of view as well as small enough in order to
minimise computational resources.  In fact, as shown by the time-delay
model (Matteucci \& Greggio, 1986), the [$\alpha$/Fe]
ratio is a powerful estimator of the duration of the SF.  Moreover,
both the predicted [Fe/H]-mass and [Z/H]-mass relationships in the
stars can be tested against the observed Colour-Magnitude Relations
(hereafter CMRs; e.g. Bower et al. 1992, Kaviraj et al 2005) and MMR. In order to clarify
this point, we recall that the O is the major contributor to the total
metallicity, therefore its abundance is a good tracer of the metal
abundance Z. Moreover, in this paper we focus on
the theoretical evolution of the $\alpha$ elements, and the O is by
far the most important.
On the other hand, the Fe abundance
is probably the most commonly used probe of the metal content in
stars, therefore it enables a quick comparison between our model
predictions and the existing literature.  

In the following [$\alpha$/Fe] ratio will always refer to 
the luminosity-weighted average over the stellar populations
that make a galaxy, unless stated otherwise. This value guarantees a robust comparison with
its observational counterpart, namely to the ``SSP-equivalent''
value inferred from the integrated spectra of elliptical
galaxies. We refer to Pipino et al. (2006) for details 
and caveats on the use of``SSP-equivalent'' abundances and abundance ratios
as proxies for the mean properties of a composite stellar populations
like an elliptical galaxy.

The star formation rate in the disc is 
\begin{equation}
\label{sfr}
\psi(t)={M_{\rm cold}\over\beta_*t_{\rm dyn}}.
\end{equation}
Here $M_{\rm cold}$ is the mass of the gas in the disc (all the gas in
the disc is cold and all the gas in the halo is hot) and $t_{\rm dyn}$
is the dynamical time (the time to complete a half rotation at the
disc half mass radius).  The parameter $\beta_*$, which determines the
efficiency of star formation has a fiducial value of $\beta_*=50$
(Guiderdoni et al 1998).

A Salpeter (1955) initial mass function (IMF) constant in time in the
range $0.1-40 M_{\odot}$ is assumed, since PM04 showed that
the majority of the photochemical properties of an elliptical galaxy
can be reproduced with this choice for the IMF.  We adopted the yields
from Iwamoto et al. (1999, and references therein) for both SNIa and
SNII.  The SNIa rate for a SSP formed at a given radius is
calculated assuming the single degenerate scenario and the Matteucci
\& Recchi (2001) Delay Time Distribution (DTD).  The convolution of
this DTD with $\psi$ (see Greggio 2005) gives the total SNIa rate, according to the following
equation:

\begin{equation}
R_{Ia}(t)=k_{\alpha} \int^{min(t, \tau_x)}_{\tau_i}{A (t-\tau) \psi(t-\tau) 
DTD(\tau) d \tau}
\label{snia}
\end{equation}
where $A(t- \tau)$ is the fraction of binary systems which give rise
to Type Ia SNe.  Here we will assume it constant (see Matteucci et
al. 2006 for a more detailed discussion).  The time $\tau$ is the
delay time defined in the range $(\tau_i, \tau_x)$ so that:

\begin{equation}
\int^{\tau_x}_{\tau_i}{DTD( \tau) d \tau}=1
\end{equation}
where $\tau_i$ is the minimum delay time for the occurrence of Type Ia
SNe, in other words the time at which the first SNe Ia start
occurring. We assume, for this new formulation of the SNIa rate that
$\tau_i$ is the lifetime of a 8$M_{\odot}$ star, while for $\tau_x$, which
is the maximum delay time, we assume the lifetime of a $0.8M_{\odot}$ star.
Finally, $k_{\alpha}$ is the number of stars per unit mass in a
stellar generation and contains the IMF.  The detailed treatment of
SNIa is a substantial improvement with respect to Paper I.
%, where the I.R.A. was adopted.

Stars are evolved between time-steps using a sub-stepping of at least
1\,Myr. During each sub-step, stars release mass and energy into the
interstellar medium. In GalICS, the enriched material released in the
late stages of stellar evolution is mixed to the cold phase, while the
energy released from supernovae is used to re-heat the cold gas and
return it to the hot phase in halo. The re-heated gas can also be
ejected from the halo if the potential is shallow enough 
(see also Paper I).  The rate of
mass loss in the supernova-driven wind that flows out of the disc is
directly proportional to the supernova rate.

%Given a mass $M$ of stars formed at some time $t_0$, we can calculate the current 
%contribution to the stellar ejection during a time-step from $t_1$ to $t_2$ as 
%\begin{equation}
%\Delta M_\star = - M \int_{m(t_1)}^{m(t_2)} [m - w(m)] \phi(m) dm
%\end{equation}
%where $m(t)$ is the mass of a star having lifetime $t$, $w(m)$ is the mass 
%of the remnant left after the star has died, and $\phi(m)$ is the IMF.   

The original formula for the chemical processing 
of the total metal content (see Paper I) has been extended to the elemental species we
deal with, so that the ejecta in the gas mass from the stellar 
population are:
\begin{equation}
\label{eqn:yield}
{\mathcal E_i}(t) = \int_{m(t)}^\infty \psi(t-t_m) ([m - w(m)]  Z_\mathrm{i,cold}(t-t_m) +  m Y_i(m))\phi(m) dm
\end{equation}
where $m(t)$ is the mass of a star having lifetime $t_m$, $w(m)$ is the mass 
of the remnant left after the star has died, and $\phi(m)$ is the IMF.   
The first term on the right hand side represents the re-introduction of 
the metals that were originally in the stars when they formed, and $Y_i(m)$ 
is the fraction of the initial stellar mass transformed via stellar 
nucleosynthesis into the element $i$. 
Throughout this work, we assume chemical homogeneity (instantaneous mixing), 
such that outflows caused by feedback processes are assumed to have the same 
metallicity as the inter-stellar medium, though in reality the material in the outflow 
is likely to be metal-enhanced (see Sec 6.3).  

\subsection{Galaxy evolution and properties}
\label{gal_ev}

The fundamental assumption is that all galaxies are born as discs at
the centre of a dark matter halo.  The transformation of disc stars
into bulge stars and of disc gas into star-bursting gas is due to
bar instabilities and mergers. Gas is never added to bulges directly
and the only gas in bulges is that coming from stellar mass loss.  The
star-bursting gas forms a young stellar population that becomes part of
the bulge stellar population when the stars have reached an age of
100$\,$Myr. We do not readjust the bulge radius when this happens.
The disc has an exponential profile, while the bulge and the star-burst
are described by a Hernquist (1990) density distribution.  The
star-burst scale is $r_{\rm burst}=\kappa r_{\rm bulge}$ with
$\kappa=0.1$.

The star formation law (Eq.~\ref{sfr}) has the same form and uses the
same efficiency parameter $\beta_*$ for all three components when we
redefine $M_{\rm cold}$ as the mass of the gas in the component and
$t_{\rm dyn}$ as the dynamical time of the component. For the
components described by a Hernquist profile, the dynamical time is
$t_{\rm dyn}=r_{0.5}/\sigma$, where $r_{0.5}$ is the half mass radius
and $\sigma$ is the velocity dispersion at the half mass radius.

The fraction of the disc mass transferred to the spheroidal component
(the bulge and the star-burst) depends on the mass ratio of the merging
galaxies.  The separation between a minor and a major merger is for a
mass ratio of 1:3.

%The solar abundances are taken from Asplund et al. (2005) ????
GalICS cannot spatially resolve galaxies, therefore we can
only predict chemical properties averaged over the galactic
radius.

\subsection{Energetics}

The SNII feedback is given by: 
\begin{equation}
\dot{m} = 2 \psi (t) \frac{\epsilon \, \eta_\mathrm{SN} E_\mathrm{SN}}{v_{esc}^2}
\label{feed}
\end{equation}
where $\epsilon$ is the efficiency of the supernova-triggered wind which 
is proportional to $v_{esc}^2$ and depends both on the porosity of the ISM (see Silk 2001 for details) 
and the mass--loading factor. This latter accounts for entrainment of interstellar gas 
by the wind and can be considered as a free parameter whose value is around 10.
Note that in the previous equation, $\eta_\mathrm{SN}$ is the number
of supernovae per unit star-forming mass, which is a prediction of the 
Initial Mass Function (IMF) chosen, and $E_\mathrm{SN}$ is the energy of a supernova, assumed 
to be $10^{51} erg$.  
 
%Equation~\ref{eqn:feed} is applied to find the fraction of 
%gas in the ISM that is lost by the galaxy and ejected in the intra halo medium. 
%We then equate the fraction of this gas that is completely ejected from the halo, 
%to the galaxy/halo escape velocity ratio. The gas ejected from the halo is added 
%to the halo reservoir where it may subsequently be accreted.

At variance with chemical evolution models as PM04, where the total Ia+II SNe feedback
is sufficient to  halt the SF, Paper I relies onto the observed correlation between AGN and velocity dispersion
(Ferrarese \& Merrit, 2000), and simply prevents gas from cooling in a halo 
which as a mass above the critical value of $\sim 10^{11}$ M$_\odot$
to quench cold gas accretion (Granato et al. 2004). % as suggested by Birnboim \& Dekel (2006).
A further halo-quenching mechanism has been implemented into GalICS by Cattaneo et al. (2008), who
showed how this further refinement leads to a better agreement between
our model predictions and SDSS observations of the luminosity function
and the colour bimodality (Baldry et al. 2004).

%\section{Results and discussion}

\section{Calibration of the model}

\subsection{Comparison to the Milky Way}
To provide a consistency check for our model, we adopt a procedure
typical of chemical evolution studies.
We first compare 
MW-like galaxies in our simulations with the known properties
of our galaxy.  Since many properties of such galaxies
have already been tested in the calibration of Paper I,
here we use the same selection criteria of MW-like galaxies
(namely $m_\mathrm{gas}/m_\mathrm{bar} = 0.10  \pm 0.05 \, , 
M_K = -23.7\,\mathrm{mag}   \pm 0.3 \mathrm{mag},
V_\mathrm{c} = 220 \,km/s  \pm 20\, km/s$  and 
requiring that the galaxy have spiral morphology)
and we show only chemical evolution predictions.
We found that $\sim$3\% of the spiral population
is made by MW-like objects in agreement with Paper I
statistics.

\begin{figure}
%\epsscale{.80}
\includegraphics[width=\linewidth]{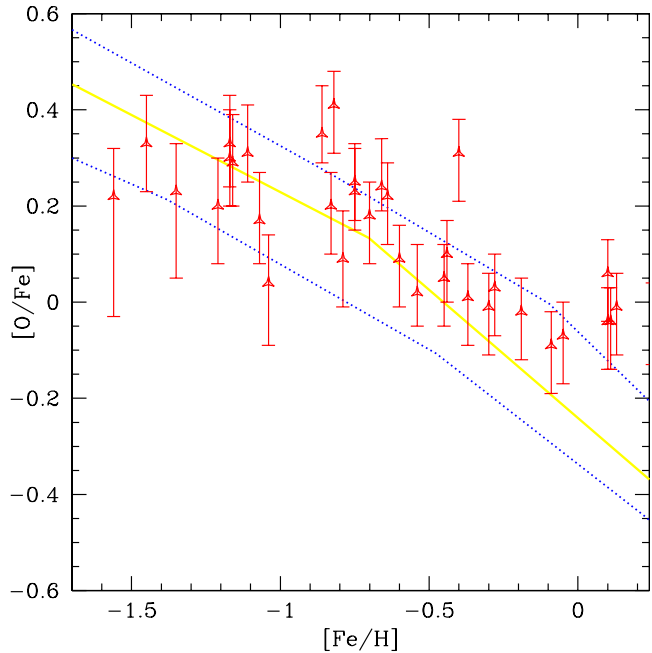}
\caption{Dotted lines: metallicity distribution of stars as
functions of [Fe/H] and [O/Fe] (3$\sigma$ contour) obtained by stacking the MW-like spirals
predicted by GalICS.  
%Dots: observational data from Fran\c ois et al. (2004).}
%\caption{
Solid line: [O/Fe] as a function of [Fe/H] as predicted 
by GalICS for a median MW-like spiral.  
Dots: observational data in the range of interest as compiled and homogenised by Fran\c cois et al. (2004).}
\label{MW}
\end{figure}

In Fig.~\ref{MW} we plot [O/Fe] ratio as a function of [Fe/H]
\emph{in the stars} of the MW-like spirals predicted by GalICS.
The dotted lines give the distribution of stars formed out of gas with a
given chemical pattern (i.e. a given [$Fe/H$] and [$O/Fe$]) as
a 3$\sigma$ contour in the [$O/Fe$]-[$Fe/H$] plane, whereas
the thick solid line give the median trend. 
The agreement with the overall trend observed in our own Galaxy 
makes us confident that the model is correctly calibrated.
Unfortunately, due to the metallicity resolution (five bins in total metallicity Z,
namely 0.001,0.004,0.008,0.02 and 0.04) of the code and to 
the fact that galaxies are identified only once their host
DM haloes are quite massive ($M_{vir} > 1.6\times 10^{11} M_{\odot}$), 
we cannot explore the region at $[Fe/H] < -1.5$.
We plan to use higher mass resolution simulations in future work, but for sake of comparison with Paper I
we restrict our analysis to the same simulation that was used in Hatton et al. (2003).

\subsection{Present-day SNIa rate}
We then verify that the model galaxies predict a present-day 
morphology dependent SNIa rate in agreement with observations.
The MW-like spirals presented above exhibit a SNIa rate of 0.09 SNuM
(i.e. specific SN explosion rate in units of $10^{10} M_{\odot}$ of stars per century).
which is in fair agreement with the observational estimates
($0.06_{-0.015}^{+0.019}$ for S0a/b and $0.14_{-0.035}^{+0.045}$ for
Sbc/d, respectively, see Mannucci et al. 2008) given the fact the GalICS does not allow a finer
morphological classification.
As shown in Fig.~\ref{snia} the vast majority of our simulated ellipticals
exhibit a present-day SNIa rate within 1$\sigma$ 
from the observational mean value given by Mannucci et al. (2008).
This test basically guarantees that, given the star formation history
of the model galaxies, we have calibrated the uncertainties
in the progenitor nature and delay time distribution of SNIa which are
incorporated in the parameter $A$ (see Eq.~\ref{snia}). 
In particular, in order to reproduce the present-day observed
SNIa rate we assume $A=0.0025$ which is the value typically
adopted in chemical evolution models of the Milky Way (see
Matteucci et al. 2006).
As a result the Fe production rate from SNIa is also calibrated.

\begin{figure}
%\epsscale{.80}
\includegraphics[width=\linewidth]{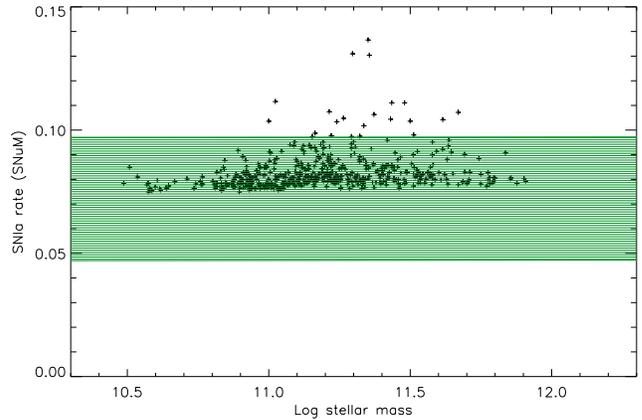}
\caption{Present-day SNIa rate in SNuM units as a function of the
galactic mass for our model ellipticals (black crosses). The hatched are
brackets the 1$\sigma$-scatter region around the observational fit (0.066
SNuM, Mannucci et al., 2008).}
\label{snia}
\end{figure}

\section{The $\sigma$- and mass-[$\alpha$/Fe] relations}
\label{main_results}

\subsection{The standard GalICS model}
\begin{figure}
%\epsscale{.80}
%\includegraphics[width=8cm,height=8cm]{mfmr.eps}
%\includegraphics[width=8cm,height=8cm]{mfsr.eps}
\includegraphics[width=\linewidth]{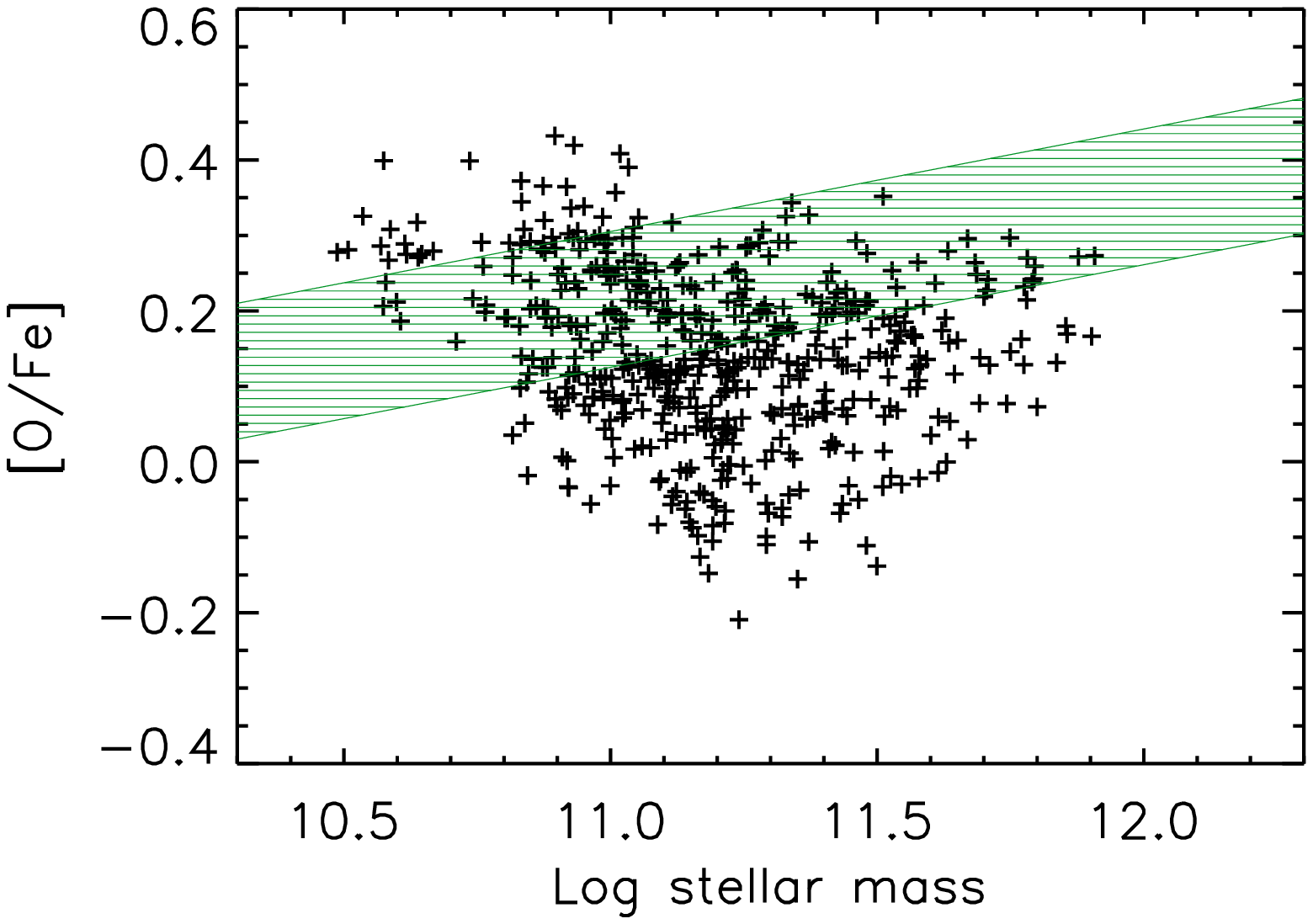}
\includegraphics[width=\linewidth]{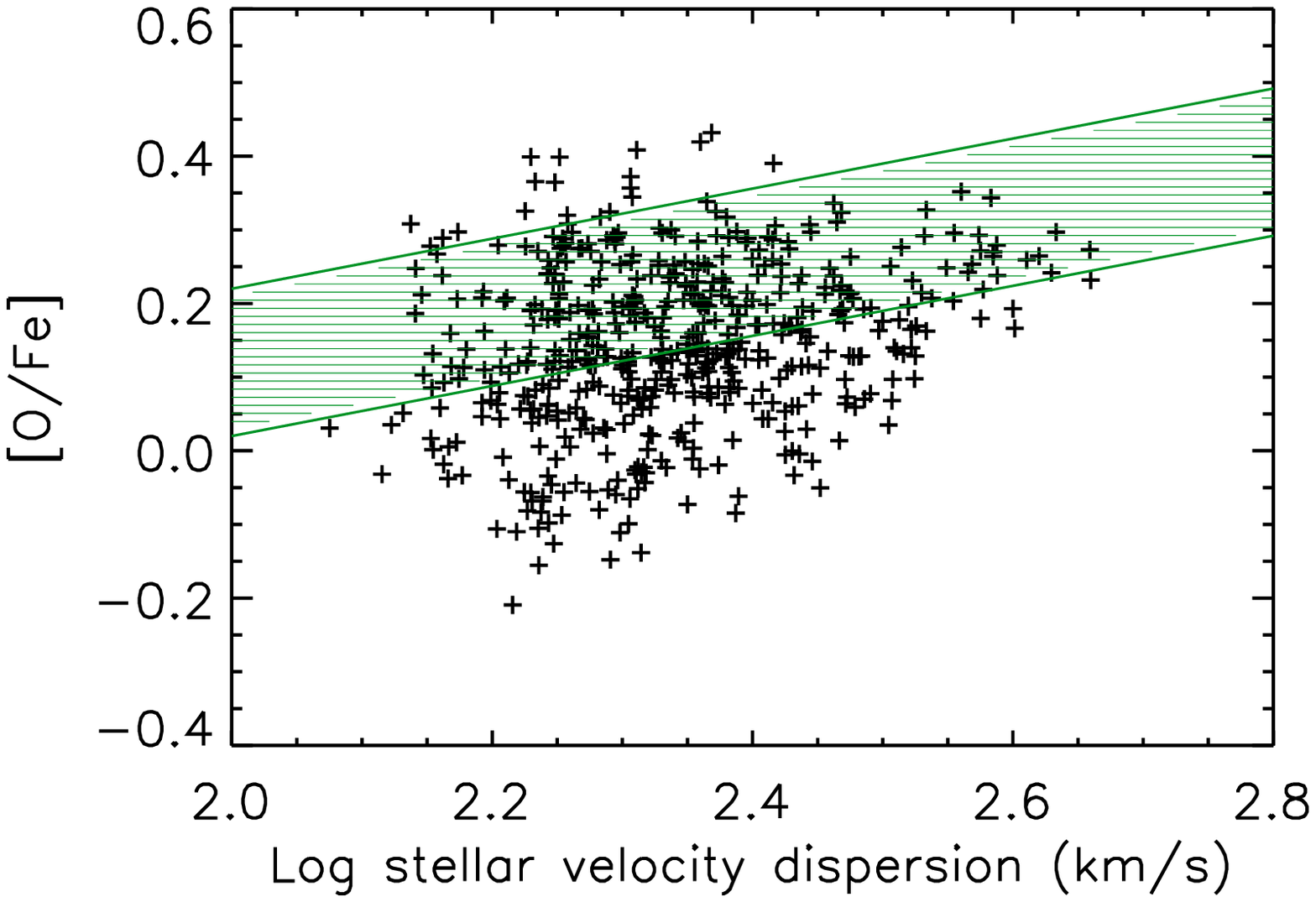}
\caption{The $\alpha/Fe$-mass and -$\sigma$ relations as predicted by GalICS for the whole 
sample of ellipticals (black points). 
The thick solid lines
encompass the 1$\sigma$-region (hatched area) around the mean trend reported by Thomas et al. (2008).
\label{mfmr}}
\end{figure}

\begin{figure}
%\epsscale{.80}
\includegraphics[width=\linewidth]{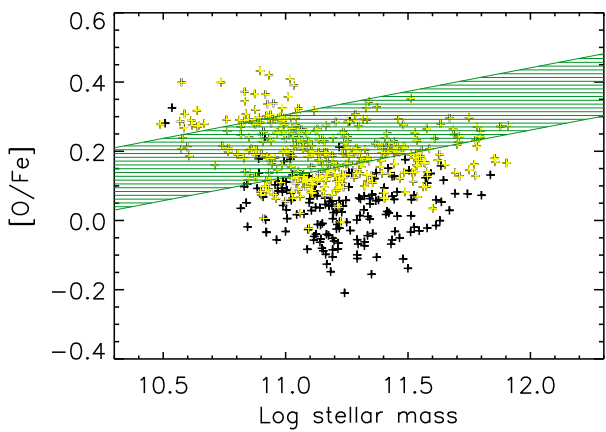}
\caption{The $\alpha/Fe$-mass relation as predicted by GalICS for the whole 
sample of ellipticals (black points). A subsample of ellipticals
older than 10 Gyr is emphasised with a lighter colour.
The thick solid lines
encompass the 1$\sigma$-region (hatched area) around the mean trend reported by Thomas et al. (2008).
\label{mfmr_bis}}
\end{figure}

From this section onwards we deal with the mean novelty
of the present work, namely the study of the predicted
$\alpha/Fe$-mass relation and its comparison to the observations.
To be consistent with observed values we present
luminosity-weighted values which take into account the disc
component (if any).
We stress, however, that the mass-weighted quantities
do not differ much from the luminosity-weighted ones
especially at the high mass end of the sample,
where SF has been suppressed at high redshift.
At variance with our previous work (PM04, PM06, PM08)
on the $\alpha$-enhancement in the very central part of the galaxies,
here we present our predictions on the [$\alpha$/Fe]
ratio in the whole galaxy (we recall that GalICS's spatial 
resolution is larger than typical galactic effective radii)
and consistently compare them to the recent observational 
estimates by Thomas et al. (2008) which pertain
to the entire galaxies. 

We also recall that observations suggest that the observed radial 
gradient slope in the [$\alpha$/Fe] has, on average, a null value (e.g. Mehlert et al. 2003).
As shown by Pipino, D'Ercole \& Matteucci (2008a), in fact, even though
the observed gradients (Carollo et al. 1993, Davies et al. 1993) suggest that
most ellipticals form outside-in, the expected strong and positive [$\alpha$/Fe] gradient 
can be affected by the metal rich gaseous flows inside the galaxy acting together
with the SFR. The net result is a gradient in the [$\alpha$/Fe] ratio
nearly flat. Hence, we can safely neglect the presence of gradients in our study.
Instead, they might affect the MMR (see discussion in Sec.~\ref{mm})

%DTstart
The results for our fiducial GalICS version are presented in Fig.~\ref{mfmr}.
In agreement with Nagashima et al. (2005), the $\alpha/Fe$ ratios do not show any correlation with mass (Fig.~\ref{mfmr}, top panel), in strong contrast with the clear positive correlation derived observationally. The simulations seem to produce decreasing [$\alpha$/Fe] ratios with increasing galaxy mass, with a slight upturn at the high-mass end. As a result the scatter is large. A linear fit of the simulation results in the [$\alpha$/Fe]-mass plane would give a flat relationship. It is interesting to note that this result does not depend on the environment in the sense that, if we restrict the regression analysis to a sub-sample of galaxies living in haloes whose mass is comparable to rich cluster of galaxies, we do not notice substantial changes in the predicted relationships.

%This somehow hints to a problem common to semi-analytical
%models, despite the different recipes adopted by different groups.
%Also, we notice that the most massive galaxies
%predicted by Nagashima et al. (2005) still have the lowest [$\alpha$/Fe] ratio 
%Therefore, our model slightly improves with respect to earlier semi-analytical models, which featured
%a decreasing trend of [$\alpha$/Fe] versus mass, completely at odds
%with the real data. 

We notice that the most massive galaxies attain a typical level of $\alpha$-enhancement
that is only 1$\sigma$ off the value suggested by the observations. This is an improvements with respect to previous results and mainly caused by the implementation of AGN feedback (see Discussion). There are two possible formation paths for these objects: 
%DTend
i) either these galaxies assemble through dry (gas-poor) mergers or ii) assemble most of their mass over very short time-scales (less than 0.5 Gyr). 
Indeed, this ensures that the pollution from SNIa is kept at
a low level and, hence, that they maintain an over-solar [$\alpha$/Fe] ratio in their stars.
However, as showed by Pipino \& Matteucci (2008), 
low-mass and highly $\alpha$-enhanced galaxies are needed if one wants
to create the most massive spheroids with a suitable $\alpha$
enhancement by means of dry-mergers. GalICS then predicts that
a small number of them, with masses $\sim 0.5-1\times 10^{11}M_{\odot}$ should
survive down to redshift zero. Unfortunately such galaxies are not observed.
In a sense one could turn the argument around and say that a robust prediction
of semi-analytic models of hierarchical galaxy formation is the 
presence of low-mass, highly [$\alpha$/Fe]-enhanced galaxies
%DTstart
at high redshift because it is the only way
in these models to build local massive ellipticals with
the observed [$\alpha$/Fe] ratios. We know from observations that such objects do not seem to exist at moderately high redshifts around $z\sim 0.4$ (Ziegler et al. 2005)%cite
%DTend

If we consider the subsample of galaxies whose luminosity-weighted
ages are larger than 10 Gyr (lighter points in fig.~\ref{mfmr_bis})
we notice that the galaxies populating the region below the observed
area in the [$\alpha$/Fe]-mass plane disappear.

We attribute this to the fact that these galaxies live
in the centre of massive haloes where both the original AGN feedback
implemented in GalICS and the halo-quenching mechanism (Cattaneo et al. 2008)
halted the cooling in the gas a long time ago.
This finding confirms the interpretation of the [$\alpha$/Fe] ratios below
the observed range as being related to a too long duration of the star formation,
as we will discuss later in the paper. However, there has been some improvement
in the agreement with observations with respect to previous models.

Similar results are obtained when plotting the [$\alpha$/Fe] as a function of 
the stellar velocity dispersion $\sigma$ (fig.~\ref{mfmr}, bottom panel).
Comparing the two panels in fig.~\ref{mfmr}, 
we notice that the scatter is somehow reduced and that the galaxies
follow a trend which is closer to the observational 
results. We can understand this as we expect the
velocity dispersion to be more correlated with the properties
of the DM host haloes, whereas the baryonic mass
is more sensitive to our modelling of feedback processes.
However, one should bear in mind that GalICS 
assumes virialization and a fixed density profile (see Paper I)
to calculate $\sigma$.
Whilst this assumptions are reasonable at z=0, 
they very likely should be revised
at high redshifts.

Moreover, we should note that the scatter is still much larger
than the observed one and that, for a given velocity
dispersion, the model galaxies tend to have on average a lower $[\alpha/Fe]$ ratio 
than the observed ones.
The latter problem can be handled in several ways (stellar yields, IMF,
feedback, SF efficiency, see Sec.~\ref{param} for details), whereas the former is
intrinsic to the model, being linked
to the stochastic nature of the merger process where
errors in the estimate of individual SF histories add up
as galaxies merge together.

%\begin{figure}
%%\epsscale{.80}
%\includegraphics[width=8cm,height=8cm]{mfsr.eps}
%\caption{The MFSR as predicted by GalICS for the whole 
%sample of ellipticals (black points). 
%The thick solid lines
%encompass the 1$\sigma$-region (hatched area) around the mean trend reported by Thomas et al. (2008).
%\label{mfsr}}
%\end{figure}

In order to better understand the origin of this scatter, we now
focus on few selected galaxies with the same mass, but very
different [O/Fe] and we study their star formation histories. 
We perform this exercise for two masses:
i) \emph{typical ellipticals} of $\sim 10^{11}M_{\odot}$;
ii) \emph{massive ellipticals} of $\sim 6\times 10^{11}M_{\odot}$.

\subsection{Typical ellipticals}

\begin{figure}
%\epsscale{.80}
\includegraphics[width=\linewidth]{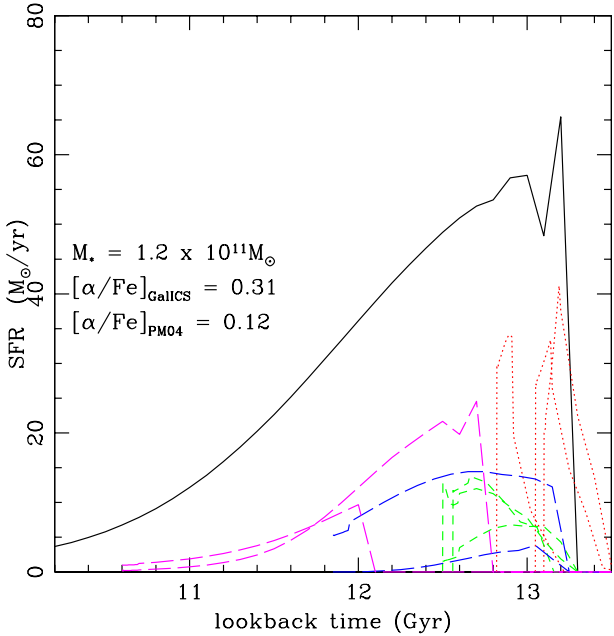}
\caption{Integral star formation history for a $\sim 10^{11}M_{\odot}$ 
galaxy with strong $\alpha$-enhancement (solid line).
The SFH inherited from the single building blocks is also shown by dotted (progenitors 
merging very early on), dashed (progenitors merging at z$\sim$4)
and long dashed (progenitors merging at z$\sim$2-3) lines, respectively.}
\label{sfh1}
\end{figure}

\begin{figure}
%\epsscale{.80}
\includegraphics[width=\linewidth]{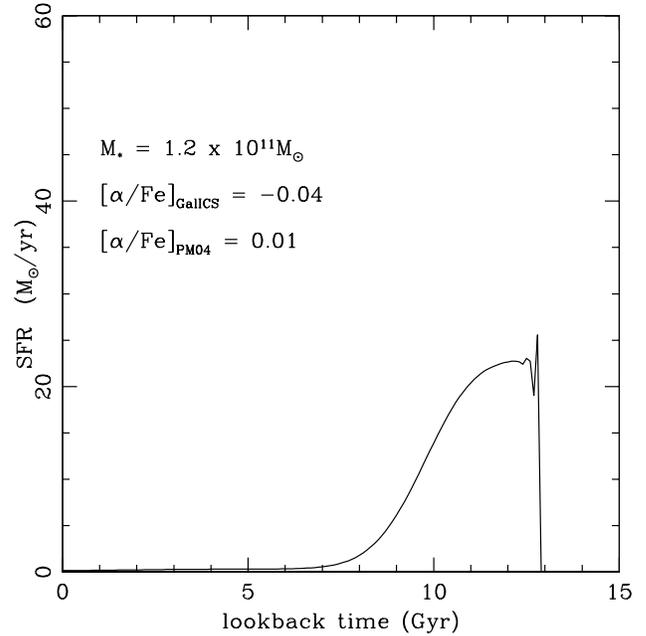}
\caption{Integral star formation history for a $\sim 10^{11}M_{\odot}$ 
galaxy with nearly solar [$\alpha$/Fe]}
\label{sfh2}
\end{figure}

%\begin{figure}
%%\epsscale{.80}
%\includegraphics[width=8cm,height=8cm]{fig7.eps}
%\caption{Integral star formation history for a $\sim 10^{11}M_{odot}$ 
%galaxy with strong $\alpha$-depletion}
%\label{sfh3}
%\end{figure}

\begin{figure}
%\epsscale{.80}
\includegraphics[width=\linewidth]{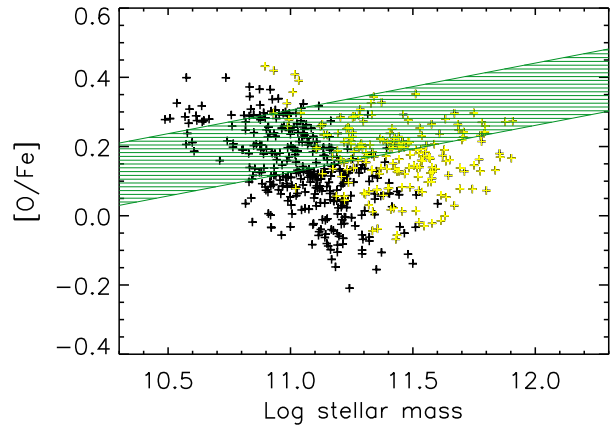}
\caption{The $\alpha/Fe$-mass relation as predicted by GalICS for the whole 
sample of ellipticals (black points). A subsample of ellipticals
which experienced at least two mergers is presented with lighter colours.
The thick solid lines
encompass the 1$\sigma$-region (hatched area) around the mean trend reported by Thomas et al. (2008).
\label{mfmr_ter}}
\end{figure}

We start the analysis in the $\sim 10^{11}M_{\odot}$ mass range, where
most of the predicted galaxies scatter outside of region of the observed
values.
A useful test bench for understanding the behaviour of such galaxies
is provided by running chemical evolution models with the same
stellar yields, IMF and the same \emph{integral} star formation rate
as the selected semi-analytic galaxies.

Indeed when we force the SFH pictured in
Fig.~\ref{sfh1} (solid line) to happen in a standard chemical evolution
model (PM04), the predicted [O/Fe] is lower by 0.2 dex than that
obtained with GalICS. This brings the galaxy from a value of [$\alpha$/Fe]=0.31 down to 
a value of [$\alpha$/Fe]=0.12 which is within the range of the observed values. 

On the other hand, for the SFH presented in Fig.~\ref{sfh2}
we have an $\alpha$-depletion with GalICS, which does
not occur in PM04. Once again this change is enough to bring 
the galaxy back within the range of observed values.

The difference here is that the galaxy whose SFH is portrayed
in Fig.~\ref{sfh1} has 9 progenitors with 4 of them
merging very early on (i.e. at $z < 4.7$), and which is
passively evolving from red-shift 2.
Looking at Fig.~\ref{sfh1} (dashed and dotted lines) we see that all the progenitors
of this galaxy actually have individual SF time-scales which are shorter
than the one that would be inferred from the mass-weighted SFH of the galaxy
itself.
This explains why the [$\alpha$/Fe] ratio calculated by GalICS is
higher than the one derived by a pure chemical evolution model
of a single object with the same mass-weighted SFH.
This is a systematic trend of hierarchical galaxy formation
when compared to pure chemical evolution models as
is shown also in Figs. ~\ref{sfh4} and~\ref{sfh5} for massive
ellipticals.
On the other hand, in the case of fig.~\ref{sfh2},  
we have only one progenitor which explains why the [$\alpha$/Fe]
ratios of GalICS and the pure chemical evolution model are in good
agreement. 
Looking at statistics with the help of Fig.~\ref{mfmr_ter} the latter case (black points)
represents 42\% of the total number of elliptical galaxies 
and it is biased towards lower mass as we would expect
since massive ellipticals are built by multiple mergers in the
hierarchical galaxy formation scenario (lighter points
in Fig.~\ref{mfmr_ter} indicate galaxies which experienced at least
two mergers.).
One might be worried that these results depend on the mass 
resolution of the N-body simulation, however 
as stated earlier in the paper our mass resolution 
is such that galaxies more massive than $2\times 10^{10}M_{\odot}$
are resolved and this mass is about a factor of 10 lower
than the mass of ellipticals considered in our analysis. This 
means that increasing the resolution would only change the number
of minor mergers (except, of course, in the very early stages
of the formation of these galaxies).
%In other words, we expect that resolution effects will systematically
%increase the [$\alpha$/Fe] ratio by a modest amount for these galaxies.
%We can have a good estimate of the maximum boost we will ever
%get from resolution by looking at how much GalICS galaxies
%with multiple progenitors differ from the prediction
%of the pure chemical evolution model; this upper value
%is 0.2 dex, which is just enough to bring back our population
%of typical ellipticals in agreement with the observed $\alpha/Fe$-mass relation.
%Therefore, we consider it very unlikely that the discrepancy
%between models and observations is simply due to resolution.
%Furthermore, this systematic effect would exacerbate the problem of low-mass
%ellipticals with [$\alpha$/Fe] ratios above the observed
%$\alpha/Fe$-mass relation.

\subsection{Massive ellipticals}

\begin{figure}
%\epsscale{.80}
\includegraphics[width=\linewidth]{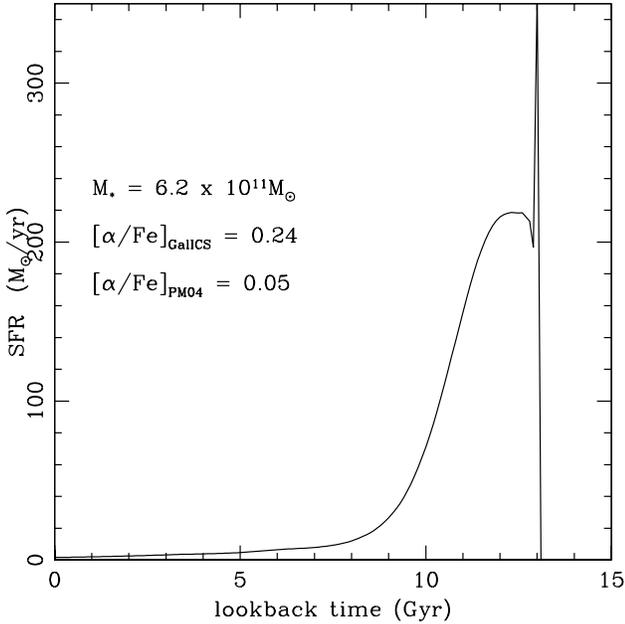}
\caption{Integral star formation history for a $\sim 6\times 10^{11}M_{\odot}$ 
galaxy with high $\alpha$-enhancement}
\label{sfh4}
\end{figure}

\begin{figure}
%\epsscale{.80}
\includegraphics[width=\linewidth]{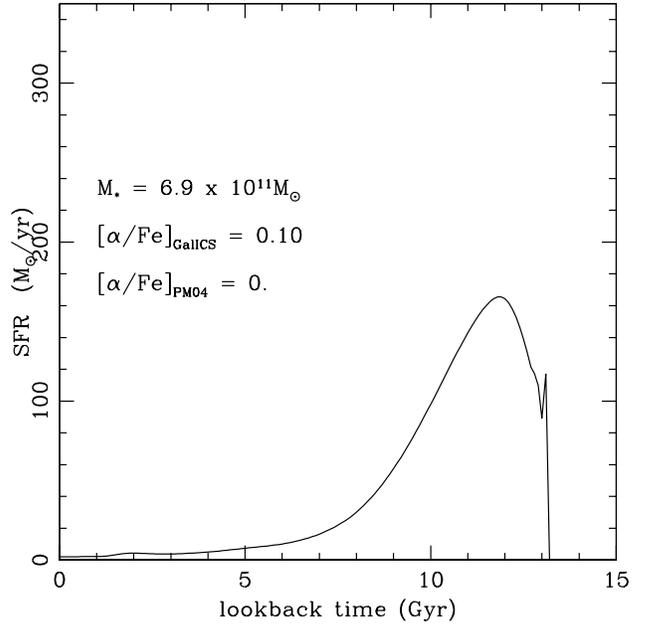}
\caption{Integral star formation history for a $\sim 6\times 10^{11}M_{\odot}$ 
galaxy with mild $\alpha$-enhancement}
\label{sfh5}
\end{figure}

If we analyse the most massive galaxies,
we confirm the findings of the previous
section (see figs.~\ref{sfh4}-~\ref{sfh5}).

In general, despite the quite high $\alpha$-enhancement, the integral
star formation history appears to be broader than
the one obtained by means of a monolithic collapse model
which would predict the same [$\alpha$/Fe] ratio.
To better explain this point, let us think to the ideal case in
which we have two progenitors with the same masses, same star formation histories (such that
their final [$\alpha$/Fe] is appropriate for their mass), but say also that their peak in the 
star formation rate is shifted by about 1 Gyr (one is younger than the other
and this difference in age cannot be detected with the standard
line-strength indices technique if this objects are more than 10 Gyr old). 
Let us also assume that these two galaxies coalesce via a dry merger later
in their evolution. The final object
still has the same [$\alpha$/Fe] of the progenitors, its mass in only doubled (therefore it
still matches the observations, given the spread in the $\alpha/Fe$-mass relation, see PM08), 
but the integral star formation history will look like the one
in Fig.~\ref{sfh4}, thus broader than the 0.5-0.7 Gyr expected by PM04 (their
model II).

We also note that in GalICS, the galaxies do not
evolve as closed box. They instead exchange metals
with the surrounding hot halo as well
as stars can be created in discs and the moved to bulges
because of instabilities. Such processes, render
the interpretation of the final [$\alpha$/Fe] ratio
on the basis of the SFH alone much more complicated.

Finally, a stacked specific (i.e. per unit stellar mass) SFH for galaxies in different
mass bins has already been presented  by Cattaneo et al. (2008)
and we do not repeat the analysis here.
With the help of the SFH presented in this section, however, we 
explained why the average duration of the SF is a factor of 3-5 longer 
(and consequently the peak value is at a factor of 3-5 lower) 
than what is required from pure chemical evolution studies
on line-strength indices analysis to reproduce the [$\alpha$/Fe] in  massive ellipticals.

%DT
%This seems to be a common feature in recent models based on 
%the hierarchical clustering which claim to have incorporated
%downsizing (e.g. De Lucia et al 2006, Kobayashi et al. 2007, Khochfar \& Ostriker, 2008),
%therefore we expect that this paper's conclusions may apply also to these models.
%To date, however, only quasi-monolithic evolution models
%are in good agreement for what concerns \emph{shape, time-scale and mean red-shift of formation}
%with those inferred by Thomas et al. (2008),
%the only difference being the sharp truncation due to the galactic wind (see PM08).

\section{The mass-metallicity relation and ages of the galaxies in the standard model}
\label{mm}

\begin{figure}
%\epsscale{.80}
\includegraphics[width=\linewidth]{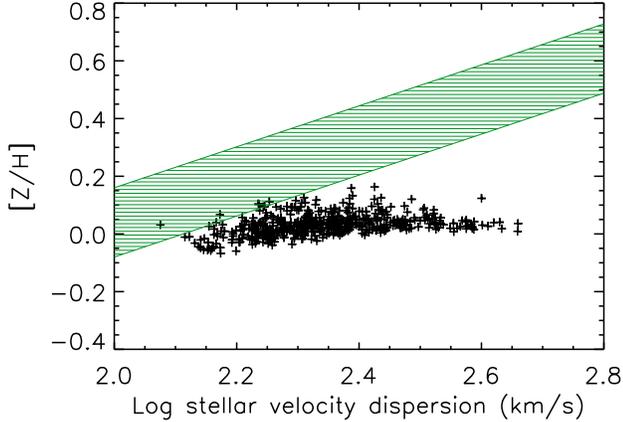}
\caption{The MMR as predicted by GalICS for the whole 
sample of ellipticals (black points). 
The thick solid lines
encompass the 1$\sigma$-region (hatched area) around the mean trend reported by Thomas et al. (2008).
Note that elliptical galaxies do exhibit quite strong 
[Z/H] gradients within one effective radius (e.g Carollo et al., 1993; Davies et al 1993)
and Thomas et al. (2008) galaxies were observed with a fixed fiber size.
Therefore it is difficult to make a meaningful comparison between our predictions
and observations as in the case of the [$\alpha$/Fe]-mass relation (see text).}
\label{mmr}
\end{figure}

In fig.~\ref{mmr} we show the predicted MMR relation 
against the data by Thomas et al. (2008).
We note that the predicted flat behaviour is unaffected
by the cuts in either mass or age as done in the Sec.~\ref{main_results}.

A failure in reproducing the MMR is expected on the basis of the preliminary analysis
by PM08 who showed that the diagnostic power of the MMR and $\alpha/Fe$-mass relation relies in the fact
that the mechanisms required to satisfy the former tend to worsen
the agreement with the latter, and vice-versa.
Only when most of the star formation process and the galactic assembly occur
at roughly the same time and the same place both relations can be fulfilled.
In the past, in fact, models were mainly aimed in reproducing the MMR,
rather than the downsizing trend, therefore they failed in the $\alpha/Fe$-mass relation.

However, elliptical galaxies do exhibit quite strong (i.e.
-0.3 dex per decade in radius)
[Z/H] and [Fe/H] gradients  within one effective radius (e.g.
Carollo et al., 1993; Davies et al 1993). %cite
Therefore it is difficult to make a thorough comparison between our predictions
and observations.
In practice, given the fixed aperture set by the SDSS fiber size,
it is likely that smaller galaxies contributed with most of their light, whereas only
the central regions (more metal rich) are observed in bigger galaxies, thus biasing
the observed slope of the MMR towards steeper values than the 
reality. However, a MMR does exist for the central regions of elliptical galaxies (e.g. Thomas
et al. 2005) with quite a similar slope to the one holding
for the entire galaxies.
Therefore, the disagreement with the observed MMR at high masses
cannot entirely explained by the fact that the GalICS spatial
resolution is not enough to take the aperture effects into account. 
We recall that, as shown by Pipino et al. (2008a, see also Sec.~\ref{gal_ev}), 
this is not a problem for [$\alpha$/Fe] ratios in the stars, because
of the interplay between internal metal flows and the star formation efficiency, which
keep the [$\alpha$/Fe] gradient flat.

%the gradient slope should correlate tightly with
%the galactic mass, and issue that it is quite controversial
%in the literature (e.g. ref).

We do not show predictions on the age-mass and colour magnitude relation.
We refer the reader to Cattaneo et al. (2008) and Kaviraj et al. (2005)
who show a remarkable agreement of the predictions made be means of GalICS 
and the latest observational results.
We only note that the predicted scatter in the predicted MMR
and (SSP-equivalent) age-mass relationships 
is comparable to the \emph{intrinsic} scatter derived by Thomas et al. (2005, 2008). 
On the other hand, the scatter in the $\alpha/Fe$-mass relation is about twice as big.

\section{Discussion}
\label{param}

\begin{figure}
%\epsscale{.80}
\includegraphics[width=\linewidth]{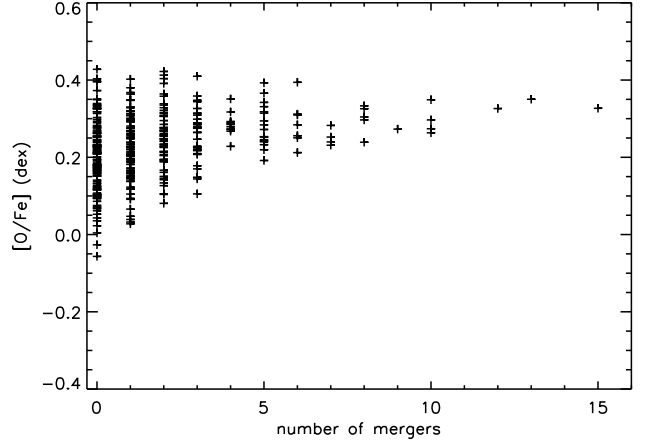}
\includegraphics[width=\linewidth]{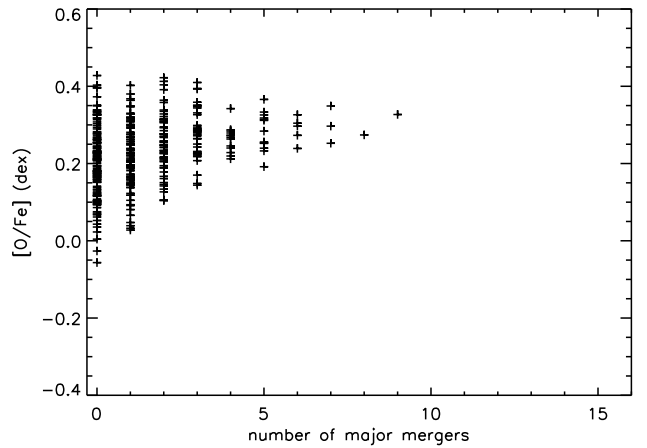}
\caption{$\alpha$-enhancement versus number total number of mergers (top panel) and number of major mergers (bottom panel)
for our model galaxies.
\label{nmerg}}
\end{figure}

\begin{figure}
%\epsscale{.80}
\includegraphics[width=\linewidth]{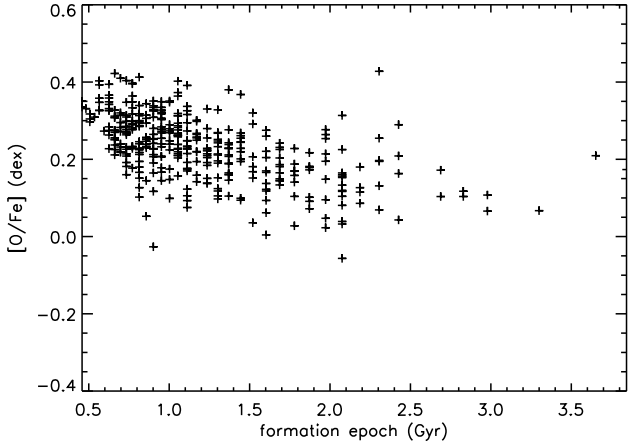}
\caption{$\alpha$-enhancement versus the epoch of the formation
of the galaxies as given by the time at which the galaxy is spotted by GalICS, 
rather than from the SSP luminosity-weighted age.
\label{epoch}}
\end{figure}

In the previous sections we showed the sensitivity of the [$\alpha$/Fe] to the integral
star formation history of the galaxy and estimated the offset in the predicted
value with respect to pure monolithic formation. Moreover,
we studied the scatter in the predicted [$\alpha$/Fe]-mass relation
in different mass ranges in order to understand the reasons for the 
persisting disagreement with observations. In this section we further discuss
this issue and try to find viable solutions to address it.

Interestingly, according to fig.~\ref{nmerg}, on average the final [$\alpha$/Fe] ratio seems
to be independent from the the total number
of mergers (top panel) as well as from the number of major mergers (bottom panel). Moreover such a relationship also holds if we consider
the number of either dry or wet mergers, and it is independent from the merger 
classification\footnote{Since gas is always present in the model galaxies,
we tested three wet merger definitions - namely when the total mass in gas divided the total baryon
mass of a newly formed galaxy exceed 0.01, 0.05 and 0.1 - and we did not
find substantial difference among them. Note that dry-mergers are
then defined by subtracting the number of wet mergers to the total for
a given galaxy.}. The model predicts a lot
of 'monolithically' (i.e. 0 mergers) behaving galaxies which exhibit a large scatter in the final [$\alpha$/Fe] ratio.
A closer inspection tell us that this fact has several reasons: 
\begin{itemize}
\item[i:] Galaxies which 
evolve from disc to bulge morphology through instabilities at late times.
\item[ii:] Galaxies in environments where the AGN-quenching
threshold occurs too late. These galaxies (along with those of case $i$)
populate the [$\alpha$/Fe]-mass plane at values lower than the
observed ones. In this case a better treatment of the feedback at galactic scales,
possibly including (SNIa-driven) winds, might represent a solution.
\item[iii:] Galaxies in environments where the AGN-quenching
threshold occurs too early (these galaxies populate the [$\alpha$/Fe]-mass plane at values higher than the
observed ones). This issue seems to be common to other models of galaxy formation
(Kimm et al., 2008). Since these galaxies are satellite
of massive haloes, one interesting (but probably not unique) possibility 
is to have them accrete some
fresh gas over a long time-scale to keep SF going and, therefore,
decrease the [$\alpha$/Fe] ratio.
\end{itemize}
 
 %DTstart
There is a trend such that galaxies with more than $\sim 5$ mergers in their formation histories always have relatively high $\alpha$/Fe ratios. As the most massive galaxies are those which underwent
the highest number of both total and wet mergers in their lifetime, these objects display relatively high $\alpha$/Fe ratios and old ages. 
%DTend
If we select again the galaxies with SSP-weighted age larger
than 10 Gyr, we find that all the galaxies with a number of mergers larger than 7 belong
to this category. (Not) surprisingly they are also the most massive 
(stellar masses larger than $5\times 10^{11}M_{\odot}$).
To be more precise, we now make use of the epoch of the
actual \emph{birth} (defined as the time at
which the galaxy is first spotted by GalICS) of the galaxy, 
rather than the SSP-weighted age. Fig.~\ref{epoch} makes clear that the higher the redshift
of the final assembly (estimated as the time at which
the last merger occurs), the higher the final [$\alpha$/Fe] ratio,
as expected from the linear regression analysis on the predicted
$\alpha/Fe$-mass relation presented earlier.
The majority of galaxies with a number of mergers in their lifetime lower than
4 (2) have formation epoch larger than 1 (2) Gyr, and have mostly
stellar masses smaller than 2 (1) $\times 10^{11}M_{\odot}$.
A confirmation of these findings comes from the pure chemical evolution study
on gaseous mergers by PM06.
They found that if a star-burst triggered by a significant accretion of pristine gas
(comparable with the mass of stars already formed - roughly similar to a major wet merger in a galaxy formation picture)
occurred at a significantly high redshift and just after the main
burst of SF, the present day photo-chemical properties of the
final elliptical galaxy match the observed ones.
Star-burst triggered at very low-redshift, instead, 
implied very low predicted final [$\alpha$/Fe] ratios,
even if the gas mass involved in the second burst was
much lower than the one converted into stars
during the main burst.

In conclusion, for the most massive spheroids, the interplay between
the peak in the merger rate and the subsequent AGN quenching of the star formation
act together in such a way that most of the star formation process and the galactic assembly occur
at roughly the same time and the same place, thus mimicking a sort
of ``monolithic'' behaviour. In other words, even though the duration
of the star formation would lead to quite low [$\alpha$/Fe] ratios
from the point of view of a pure chemical evolution model,
the fact that it had happened in several sub-units makes the final [$\alpha$/Fe] ratios
higher and in better agreement with observations.
Intermediate and small objects, instead, do not have a quenching mechanism
acting directly at their scales which can self-regulate the duration
of the star formation. Therefore they end up having either too high
or too low [$\alpha$/Fe] ratios.

\begin{figure}
%\epsscale{.80}
\includegraphics[width=\linewidth]{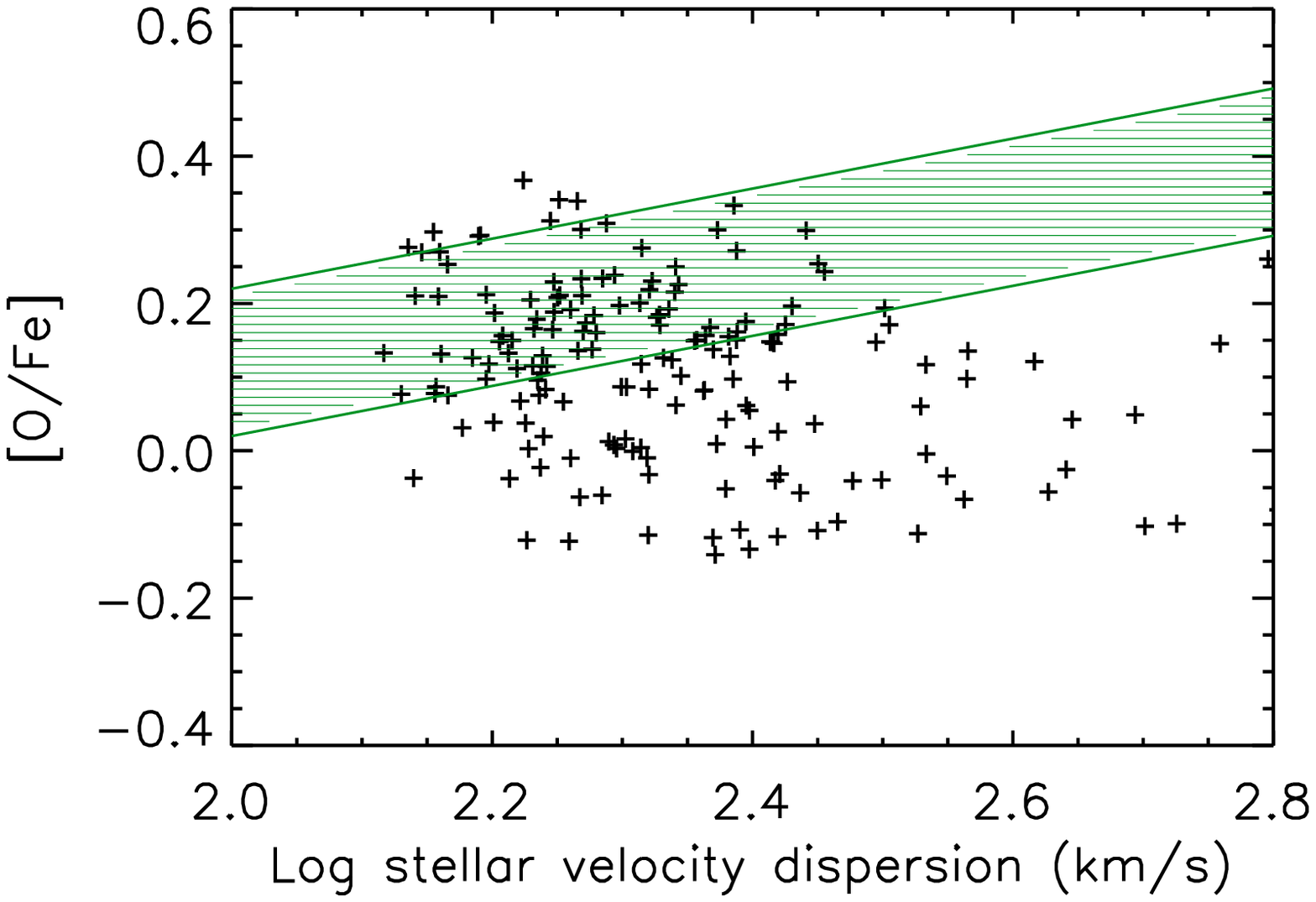}
\includegraphics[width=\linewidth]{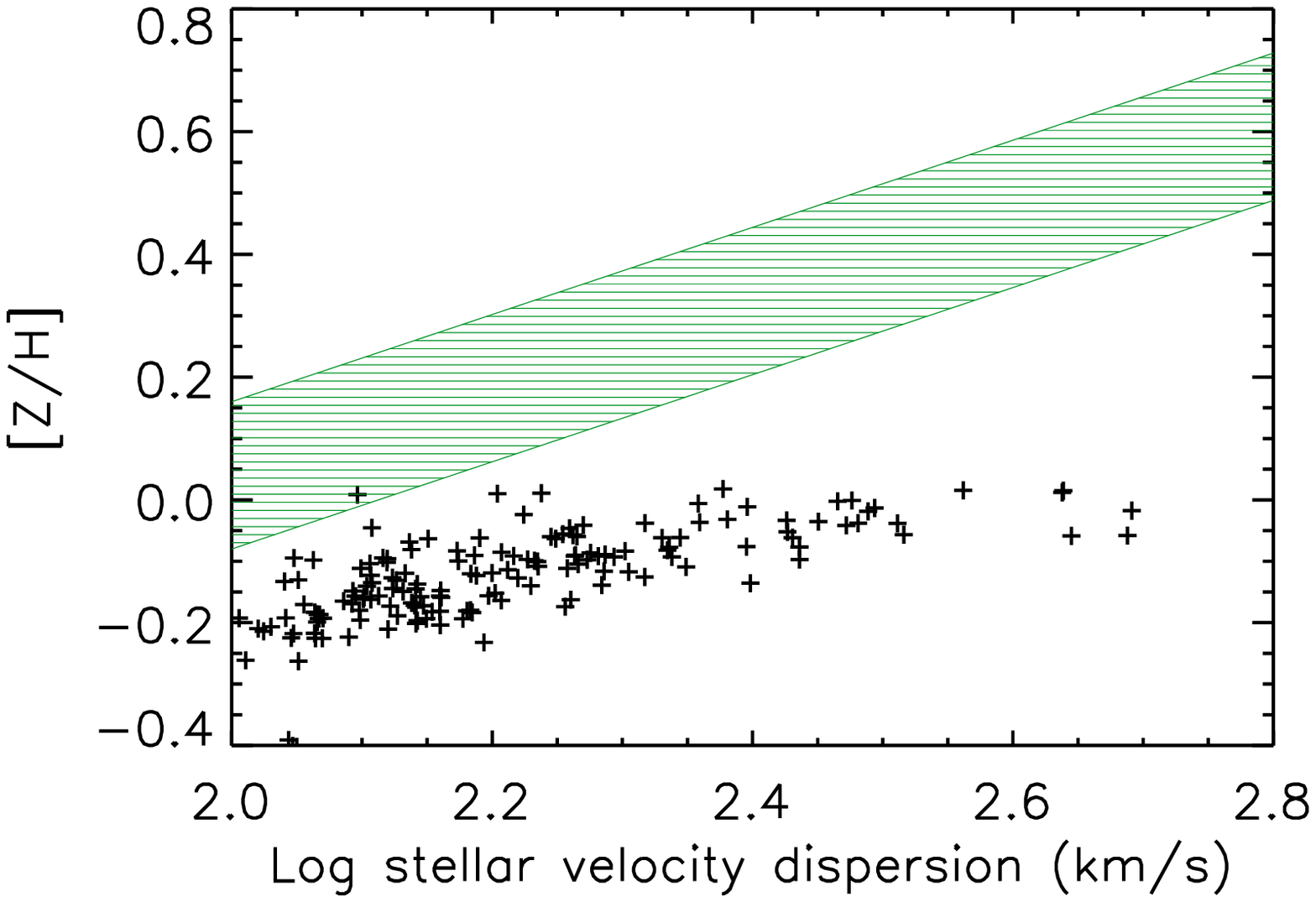}
\caption{The $\alpha/Fe$-mass (top panel) and the MMR (bottom panel) relations as predicted by GalICS for the whole 
sample of ellipticals (black points). In this run
the AGN feedback has been switched off.
The thick solid lines
encompass the 1$\sigma$-region (hatched area) around the mean trend reported by Thomas et al. (2008).
Note that in this case we have fewer ellipticals than in fig.~\ref{mfmr}
simply because gas is allowed to cool onto a disc at the centre of massive
DM haloes, which leads us to classify more central galaxies as disc-dominated spirals.
\label{mfsr_noAGN}}
\end{figure}

In the remainder to this section we discuss the main ingredients that might be modified
in order to improve the predicted [$\alpha/Fe$]-mass relation.
%We note in passing that most of the parameters in GalICS hardly modify the SFH, and thus, 
%the SNIa explosion rate. 

\subsection{AGN feedback}
We start presenting the key factor that permitted substantial improvement.
If we switch off the SMBH heating of the intra-halo gas,
we predict that the more massive galaxies are younger than the less massive
ones - with a typical age of 6 Gyr (at variance with
the observational results) -  and that they are strongly $\alpha$-depleted. 
As can be seen in fig.~\ref{mfsr_noAGN} (top panel), we basically confirm
with a more self-consistent model the results that Thomas (1999)
obtained: the natural prediction of a bottom-up
hierarchical growth of the galaxies leads to a $\alpha/Fe$-mass relation which
has a negative slope.

%DTstart
Therefore some kind of SF quenching is needed. The recipes commonly adopted in the literature (see Introduction) and in the present work assume that
only the most massive galaxies are affected by SMBH feedback. This helps to predict higher [$\alpha$/Fe] ratios and relaxes, but does not solve yet, the problem at the high mass. As discussed above, however, little attention is paid to what becomes of the less massive objects which are the building blocks of the relation and responsible for the bulk of the scatter. Suppression of star formation at low and intermediate masses is required in the models. In fact, recent observations show that AGN feedback appears to be present also in this mass range of the early-type galaxy population (Schawinski et al 2007)%cite.
%DTend

The bottom panel of fig.~\ref{mfsr_noAGN} show us that the predicted MMR slope is in better agreement
with the data, although offset downwards in line with the above discussion (see Sec.5) and PM08's analysis.

\subsection{Stellar yields} 
A change in the stellar yields will introduce
a systematic offset of a few tenths of a dex in the model predictions (see Thomas
et al., 1999, PM04), hence it might let all the massive
galaxies lie within the observational boundaries
for a suitable choice of the stellar nucleosynthesis is done\footnote{For
instance by: i) extending the upper mass limit of the IMF to 100 $M_{\odot}$,
ii) neglecting the ejecta of the stars in the mass range 8-11 $M_{\odot}$
(whose O/Fe is slightly sub-solar); iii) by setting the SNIa rate
to the lowest value permitted by observations.}.
However, being only an offset, this change will not
modify the slope of the predicted $\alpha/Fe$-mass relation and will exacerbate the
problems for the low mass galaxies.
%DTstart
But most importantly, the successful calibration of our model with element ratios observed in Milky Way stars does not allow significant modifications of the underlying stellar yields.
%DTend
%Nonetheless, the role of the stellar yields must be kept
%in mind because it is fundamental if we want to extract
%information from the zero point of the $\alpha/Fe$-mass relation.

\subsection{IMF}

We do not test other IMFs, since
a Salpeter IMF is a valid assumption for explaining most of the properties
of early-type galaxies (see Renzini 2005).
Furthermore, it has been already shown by Nagashima et al. (2005) that a change in the IMF
is not enough (in the context of hierarchical mergers)
in reproducing the correct $\alpha/Fe$-mass relation. 
We expect that a flattening of the IMF
regardless of the galaxy mass leads to a overall shift towards
higher values of the predicted [$\alpha$/Fe] ratio, but it does not
affect the slope of the $\alpha/Fe$-mass relation unless one finds a good reason to
make the IMF flatter as the galactic mass increase.
%DTstart
We will investigate this possibility in future work. This solution appears contrived, however, because it implies that the single building block should know in advance its destiny in order to self-assign a suitable IMF.
%DTend

\subsection{Feedback from SNe and its efficiency}
In Eq. (7) $\epsilon^{-1}$ represents the efficiency of mass--loading during 
the triggering of a galactic wind by SNII explosions.  
Decreasing $\epsilon$ produces more feedback, heating more cold 
gas, ejecting more hot gas from halos, and thus reducing the amount of 
gas that can potentially form stars. In this case
we find that the predicted stellar masses are smaller
than the fiducial case. The galaxies look slightly 
more $\alpha$-enhanced as expected since the SF process is
strongly-disfavored by the SNe explosions.
However, this is not a viable solution for the $\alpha/Fe$-mass relation problem,
since the high mass--loading also implies a very low
metal content in the stars. The predicted MMR
is offset downwards by at least 0.5 dex from the observational one.
On the other hand, if we switch the SN feedback off, we
tend to slightly worsen the $\alpha/Fe$-mass relation, whereas the agreement
for the MMR improves.

Following the above line of thoughts, we further modify GalICS
by introducing SNIa contribution into Eq.~\ref{feed}.
Since $\alpha$ elements and Fe are still ejected at the same rate,
this change has the same effects of decreasing $\epsilon$\footnote{With
the obvious difference that we can have ejection of matter
also when the SF is zero, because of the nature of SNIa progenitors.}.
A 0.1 dex increase in the final $[\alpha/Fe]$ ratios can be achieved
when a differential wind is invoked, namely if we assume that twice
more Fe than O can be ejected in the hot phase due to
SNIa explosions.
Again, given the nature of such a mechanism, neither the slope
of the predicted $\alpha/Fe$-mass relation can be steepened nor its scatter
reduced. Further investigation will tell us if a change
in the SNe feedback, namely by allowing them to quench the star formation
as in monolithic models (e.g. Pipino et al., 2008b), might be the required galactic-scale
source of feedback.

\subsection{Star formation efficiency}

We identify the SF recipe as one of the prescriptions
where one can improve upon. In fact, Pipino et al. (2008b)
started from the heuristic approach of PM04,
who required the SF efficiency to increase as a function
of galactic mass, and showed that the $\alpha/Fe$-mass relation can be explained 
by implementing a physically motivated value for the SF efficiency.
To explain the higher star formation efficiency in the most massive galaxies,
they appeal to massive black holes-triggered SF:
a short ($10^6-10^7$ yr) super-Eddington phase can 
provide the accelerated triggering of associated star formation.
The SMBH grows mostly in the initial super-Eddington phase while 
most of the spheroid stars grow during the succeeding Eddington phase, until
the SN-driven wind quenches  SF. 
According to Pipino et al. (2008b) models, the galaxy is fully assembled on a time-scale of 0.3-0.5 Gyr.
This time-scale is long enough, however, to allow the SMBH to complete its growth
in order to reproduce the Magorrian  relation.
The fact that GalICS already turns gas into stars
at the maximum possible rate during the merger-induced star-burst phase
and its quite low mass/space resolution hamper us from  
a direct implementation of the above recipe. 
Moreover, the fact that stars born the disc and can be transferred to the bulge
of the same galaxy because of instabilities, is a possibility
not taken into account in Silk (2005).
This scenario will be tested in the forthcoming version of GalICS.

\section{Conclusions}

We implemented a detailed treatment for the chemical evolution of H, He, O
and Fe in GalICS, a semi-analytical model for galaxy formation which
successfully reproduces basic low- and high-redshift galaxy properties.
The contribution of supernovae (both type
Ia and II) as well as low- and intermediate-mass stars to 
chemical feedback are taken into account. The model predictions are
compared to the most recent observational results by Thomas et al. (2008).
We find that this chemically improved GalICS does not produce the observed mass- and $\sigma$-[$\alpha$/Fe] relations. The slope is too shallow and scatter too large, in particular in the low and intermediate mass range. The model shows significant improvement at the highest masses and velocity dispersions, where the predicted [$\alpha$/Fe] ratios are now marginally consistent with observed values. 
Moreover, an excess of low-mass ellipticals
with too high a [$\alpha$/Fe] ratio is predicted.
We show that this result comes from the implementation of AGN (plus halo) quenching of the star formation in massive haloes. 

A thorough exploration of the parameter space shows
that the failure of reproducing the mass- and $\sigma$-[$\alpha$/Fe] relations can partly be attributed to the way in which star formation and feedback are currently modelled. The merger process is responsible for a part of the scatter. We suggest that the next generation of semi-analytical model should feature feedback (either stellar of from AGN) mechanisms linked to single galaxies and not only to the halo, especially in the low and intermediate mass range.

Furthermore, a drawback of the 
is the fact that the MMR cannot be fit simultaneously.
Both effects can be explained by the fact that
the model is still lacking a sort of \emph{monolithic}
formation for all its spheroids, which is needed in order
to reproduce the $\alpha/Fe$-mass relation and the MMR at the same time.
The scatter is somehow intrinsic to the merger history,
thus calling for further modification
of the baryons behaviour with respect to the CDM.
In other words we envisage a lack of a self-regulating
mechanisms which acts on a galactic scale and counterbalances
to some extent the random nature of the merger trees.

As expected from chemical evolution studies,
is the shape of the SFH which sets the final [$\alpha$/Fe]:
a galaxy with a shorter duration of the SFH (summed
over all the progenitors) will have a higher [$\alpha$/Fe]
than a galaxy with a longer one, even if the latter had less mergers.
Moreover the [$\alpha$/Fe] achieved by the galaxies
are in general 0.1-0.3 higher than what expected
by feeding the integral SFH in a pure chemical evolution model.
This happens because in GalICS galaxies do not
evolve as closed boxes. They instead exchange metals
with the surrounding hot halo, undergo dry-mergers, as well
as stars can be created in discs and the moved to bulges
because of mergers or instabilities.

In order to understand such a difference and
to find viable solutions
we tested the effect of several model parameters.
Among those, we emphasise that
an increase in the star formation efficiency 
and Fe-enhanced winds driven by the SNIa activity
might play a role in removing galaxies
with too a low [$\alpha$/Fe] ratio.
However, given the way they act on the galaxy
evolution, the cannot be effective in either making
the slope of the predicted $\alpha/Fe$-mass relation steeper or 
in reducing its scatter.
In particular, it seems hard to remove the low-mass
galaxy too $\alpha$-enhanced.

%\acknowledgments
\section*{Acknowledgments} 
AP acknowledges useful discussions with A. Cattaneo.

\clearpage

\end{document}